\documentclass[fleqn,usenatbib]{mnras}

\usepackage{newtxtext,newtxmath}

\usepackage[T1]{fontenc}

\DeclareRobustCommand{\VAN}[3]{#2}
\let\VANthebibliography\thebibliography
\def\thebibliography{\DeclareRobustCommand{\VAN}[3]{##3}\VANthebibliography}

\usepackage{graphicx}	
\usepackage{amsmath}	

\title[Constraining ULA by $\it{CSST}$]{Constraining Ultralight Axions with CSST Weak Gravitational Lensing and Galaxy Clustering Photometric Surveys}

\author[H. Lin et al.]{
Hengjie Lin$^{1,2}$,
Furen Deng$^{1,2}$,
Yan Gong$^{1,2,3}$\thanks{E-mail: gongyan@bao.ac.cn},
Xuelei Chen$^{1,2,4,5}$
\\
$^{1}$National Astronomical Observatories, Chinese Academy of Sciences, 20A Datun Road, Beijing 100012,
China\\
$^{2}$School of Astronomy and Space Sciences, University of Chinese Academy of Sciences, Beijing 100049, China\\
$^{3}$Science Center for China Space Station Telescope, National Astronomical Observatories, Chinese Academy of Sciences, \\20A Datun Road, Beijing 100101, China\\
$^{4}$Department of Physics, College of Sciences, Northeastern University, Shenyang 110819, China\\
$^{5}$Centre for High Energy Physics, Peking University, Beijing 100871, China
}

\date{Accepted XXX. Received YYY; in original form ZZZ}

\pubyear{2015}

\begin{document}
\label{firstpage}
\pagerange{\pageref{firstpage}--\pageref{lastpage}}
\maketitle

\begin{abstract}
Ultralight axion (ULA) can be one of the potential candidates for dark matter. The extremely low mass of the ULA can lead to a de Broglie wavelength the size of galaxies which results in a suppression of the growth of structure on small scales. In this work, we forecast the constraint on the ULA particle mass $m_{\text{a}}$ and relative fraction to dark matter $f_{\text{a}} = \Omega_{\text{a}}/\Omega_{\text{d}}$ for the forthcoming Stage IV space-based optical survey equipment $\it{CSST}$ (China Space Station Telescope). We focus on the $\it{CSST}$ cosmic shear and galaxy clustering  photometric surveys, and forecast the measurements of shear, galaxy, and galaxy-galaxy lensing power spectra (i.e. 3$\times$2pt). The effects of neutrino, baryonic feedback, and uncertainties of intrinsic alignment, shear calibration, galaxy bias, and photometric redshift are also included in the analysis. After performing a joint constraint on all the cosmological and systematical parameters based on the simulated data from the theoretical prediction, we obtain a lower limit of the ULA particle mass $\text{log}_{10}(m_{\text{a}}/\text{eV}) \geqslant -22.5$ and an upper limit of the ULA fraction $f_{\text{a}} \leqslant 0.83$ at 95\% confidence level, and $\text{log}_{10}(m_{\text{a}}/\text{eV}) \geqslant -21.9$ with $f_{\text{a}} \leqslant 0.77$ when ignoring the baryonic feedback. We find that the CSST photometric surveys can improve the constraint on the ULA mass by about one order of magnitude, compared to the current constraints using the same kind of observational data.
\end{abstract}

\begin{keywords}
cosmological parameters -- large-scale structure of Universe -- dark matter.
\end{keywords}



\section{Introduction}

Based on the current cosmological observations, about $95\%$ of our Universe is dark and composed of the so-called dark matter and dark energy. Although the existence of dark matter and dark energy has been supported by various observational data, such as cosmic microwave background (CMB) \citep[e.g.][]{hinshaw2013wmap, Planck2018-I}, Type Ia supernovae \citep[e.g.][]{scolnic2018IaSNe, Riess-2021IaSNe} and baryon acoustic oscillations (BAO) \citep[e.g.][]{alam2021eBOSS}, the nature of dark matter and dark energy is still puzzling.
For dark matter, the most famous and popular model is the cold dark matter (CDM) model, which can successfully provide excellent predictions compared to the observational data. However, there are still some problematic issues for this model, e.g. the "missing satellites" problem \citep[e.g.][]{Moore-1999, Klypin-1999}, the "core-cusp" problem \citep[e.g.][]{Dutton-2019, Read-2018, Genina-2018}, the "too-big-to-fail" problem \citep{Boylan-Kolchin-2011}, and the "$S_{8}$ tension" \citep{Cosmology-intertwined}. It is still not clear that, whether better understanding of astrophysical process or new dark matter model is needed to solve these problems.

In order to solve the problem of the CDM model, various alternative dark matter models have been proposed, such as warm dark matter \citep[WDM;][]{Bode-2001, Abazajian-2006}, which has non-negligible thermal motions with a mass of a few keV, self-interacting dark matter \citep[SIDM;][]{Spergel-2000, Wandelt-2001}, which has strong interactions in contrast to the standard CDM model, and the ultralight axion (ULA) from string theory compactification, produced by field misalignment before inflation \citep{Arvanitaki-2010}. 

The string theory can accommodate a very large number of the ULAs with a broad range of masses, from about $m_{\rm a}=10^{-33}$ eV to $10^{-18}$ eV, and the ULA can be a candidate of dark matter particle or even dark energy. An interesting property of the ULA is that, the equation of state of the ULA will transform with the evolution of the Universe. At early times, when the Hubble parameter $H \gg m_{\text{a}}$, the ULA equation of state is $w_{\rm a} \simeq -1$, and the ULA behaves like dark energy. At late times, when $H < m_{\text{a}}$, the equation of state oscillates around $w_{\rm a} = 0$ \citep{Marsh-Axion-Cosmology}, and the ULA acts as dark matter. As a pseudo bosonic scalar field, the ULA field behaves as nonrelativistic matter in the form of a Bose-Einstein condensate (BEC) at late times. Due to its extremely small particle mass, the corresponding de Broglie wavelength can even achieve the size of galaxies. This leads to an effective potential which opposes the gravitational force, called "quantum pressure" \citep{Tsujikawa-2021}, and it can offer a promising solution to solve the CDM problems. For example, the "quantum pressure" of the ULA field can suppress the gravitational collapse on small scales, and this can result in the reduction of the number of dwarf galaxies \citep{Woo-2009}, which can solve the "missing satellites" problem. On the other hand, the quantum pressure can support a soliton core inside a dark matter halo \citep[e.g.][]{Hu-2000, Schive-2014, Marsh-2014}, and it can explain the "core-cusp" problem. In addition, since sufficiently light axions with masses around $10^{-24}$ eV do not cluster below scales of a few Mpc, it could be a potential solution of the $S_{8}$ tension \citep{Rogers-2023}.

The current constraints on the ULA are mainly based on astronomical observations. On the largest observational scales, by using the CMB data from $\it{Planck}$, \cite{Hlozek-2018} rules out axion contributing all dark matter in the mass range $10^{-33} \leqslant m_{\text{a}} \leqslant 10^{-24}$ eV, and obtains a percent-level bounds on the fractional contribution to dark matter. By using galaxy clustering statistics from the Baryon Oscillation Spectroscopic Survey (BOSS), \cite{lague-2022} also finds a upper limit on the fractional contribution from about $1\% \sim 10\%$ for the mass range $10^{-33} \leqslant m_{\text{a}} \leqslant 10^{-24}$ eV,  which is an independent probe apart from CMB observation. On the other hand, in order to explore the higher mass range, we need to investigate smaller scales. For example, the observation of Lyman-$\alpha$ forest obtains a constraint on ULA mass scale around $m_{\text{a}} \sim 10^{-22}$ eV \citep[e.g.][]{Irsic-2017, Kobayashi-2017, Armengaud-2017}, and the strongest bound only allows axions to contribute all dark matter if $m_{\text{a}} > 2\times 10^{-20}$ eV \citep{Rogers-2021}. 

Besides the large-scale structure observations, there are lots of efforts to constrain the ULA using galactic structure or dynamics. Since the ULA can form a soliton core inside the dark matter halos, it can be explored by investigating the inner density profile of the Milky Way \citep{Bar-2018, DeMartino-2020}, or the Milky Way dwarf satellites \citep{Calabrese-2016, Chen-2017, Safarzadeh-2020, Broadhurst-2020}. Moreover, the ULAs also influence the subhalo mass function, since it can  suppress matter collapse on small scales, the abundance of low-mass halos would be lower than the CDM case.  \cite{Nadler-2021} presents a method to constrain the ULA by investigating the abundance of the observed Milky Way satellites. These methods above can explore the ULA mass range around $10^{-22} \sim 10^{-21}$ eV, though there are still large uncertainties because of the complex and unclear astrophysical process at small scales, such as baryonic feedback, tidal stripping and so on. So we need other independent probes to confirm these constraint results.

Cosmic shear survey is a powerful probe for the cosmic structure \citep{Kaiser-1992, Van-Waerbeke-2000}, and it can provide an effective way to explore the mass range of the ULA as dark matter. \cite{Dentler-2022} obtains a lower limit of $m_{\text{a}} > 10^{-23}$ eV for axion contribute to all dark matter, by combining $\it{Planck}$ and Dark Energy Survey (DES) Year 1 shear maesurement data. \cite{Kunkel-2022} investigates the constraint using power spectra, bispectra and trispectra for a $\it Euclid$-like weak lensing survey. They find that their method is able to distinguish the ULA and CDM models up to $m_{\text{a}} = 10^{-22}$ eV, which implies that the next-generation Stage-IV surveys can provide strong constraint on the ULA.

In this work, we will forecast the constraint on the ULA with the upcoming China Space Station Telescope ({\it CSST}) \citep{zhan2011csst, zhan2018csst, zhan2021csst, Gong-CSST-2019} photometric surveys, by combining cosmic shear, galaxy angular clustering and galaxy-galaxy lensing surveys, i.e. 3$\times$2pt. The {\it CSST} is a Stage-IV 2-m space telescope operating in the same orbit with the China Space Station, and is planned to launch around 2024. The main scientific goals of {\it CSST} are to study the property of dark energy and dark matter, and the large-scale structure of the Universe, which requires large survey area and deep survey depth. The {\it CSST} wide survey plans to cover 17500 deg$^{2}$ sky area in about 10 years with the field of view 1.1 deg$^{2}$. It can cover 255-1000 nm by using seven photometric bands (i.e. {\it NUV}, {\it u}, {\it g}, {\it r}, {\it i}, {\it z}) and three spectroscopic bands (i.e. {\it GU}, {\it GV}, and {\it GI}), that allows us to receive photons from near-UV to near-IR. Combined with multiple backend scientific equipments, we are allowed to collect photometric imaging and slitless spectroscopic data in the mean time. And the magnitude limit of the {\it CSST} photometric survey can reach $i \simeq 26$ AB mag for $5\sigma$ point source detection. In this study, we will generate the mock data of the $\it CSST$ photometric survey, and consider both the ULA mass $m_{\text{a}}$ and its fractional dark matter contribution $f_{\text{a}} = \Omega_{\text{a}}/\Omega_{\text{d}}$ in the constraint process, where $\Omega_{\text{a}}$ and $\Omega_{\text{d}}$ are the ULA and total dark matter energy density parameters, respectively.  The effects of neutrino, baryonic feedback, and systematical uncertainties, such as intrinsic alignment, shear calibration, galaxy bias, and photometric redshift (photo-$z$), also are included.

The paper is organized as follows: in Section \ref{sec:ULA}, we describe the ULA model we consider; in Section \ref{sec:Mock-data}, we presents the details of generating the 3$\times$2 pt mock data including relevant systematic, and describe the model fitting method; in Section \ref{sec:Results}, we discuss the constraint results. Finally, in Section \ref{sec:Conclusion}, we summarize the conclusion and give relevant discussions.

\section{Ultralight Axion Model}\label{sec:ULA}
\subsection{Basic Axion Physics}\label{subsec:ULA-physics}

Axion is initially motivated as a solution to solve the strong charge parity problem in quantum chromodynamics (QCD), and the mass of axion particle is related to the Peccei-Quinn (PQ) symmetry-breaking scale, $F_{\text{ax}}$. In order to avoid the axion relic density too high that will lead to an overclose of the Universe, QCD axions must obey the constraint $F_{\text{ax}} \lesssim 10^{12}$ GeV or $m_{\text{a}} \gtrsim 10^{-6}$ eV \citep[e.g.][]{QCD_axion_1, QCD_axion_2, Abbott-1983, Dine-1983}. 

On the other hand, axions also appear naturally in string theory, and in this case, they can have many orders of magnitude lighter mass than the QCD axions\citep[e.g.][]{Arvanitaki-2010, Marsh-Axion-Cosmology, Hui-2017, Mehta-2021}. In this scenario, axions are produced by a process called vacuum realignment, and leave a relic density given by \citep[]{Tanabashi-2018}
\begin{equation}
\begin{split}
    \Omega_{\rm a} h^{2}=0.12\left(\frac{m_{\text{a}}}{4.7 \times 10^{-19} \mathrm{eV}}\right)^{1 / 2}\left(\frac{f_{\text{a}}}{10^{16} \mathrm{GeV}}\right)^{2}\left(\frac{\Omega_{\rm m} h^{2}}{0.15}\right)^{3 / 4}\\
    \times\left(\frac{3.4 \times 10^{3}}{1+z_{\mathrm{eq}}}\right)^{3 / 4} \theta_{\rm I}^{2}.
\label{eq:relic-density}
\end{split}
\end{equation}
Here $f_{\text{a}} = \Omega_{\text{a}}/\Omega_{\text{d}}$ is the axion fraction, where $\Omega_{\text{d}}$ is the total dark matter density parameter including both axions and ordinary CDM, $h$ is the reduced Hubble constant today in units of $100\,\rm km\, s^{-1}Mpc^{-1}$, $z_{\rm eq}$ is the redshift of matter-radiation equality, and $\theta_{\rm I}$ is the initial misalignment angle. In this scenario, axions act as a coherent scalar field with extremely light mass, i.e. $10^{-33} \lesssim m_{\text{a}} \lesssim 10^{-18}$, and this is the so-called ultralight axion,  which can play a role of the "fuzzy" cold dark matter.

\begin{figure*}
    \centering
    \includegraphics[scale = 0.5]{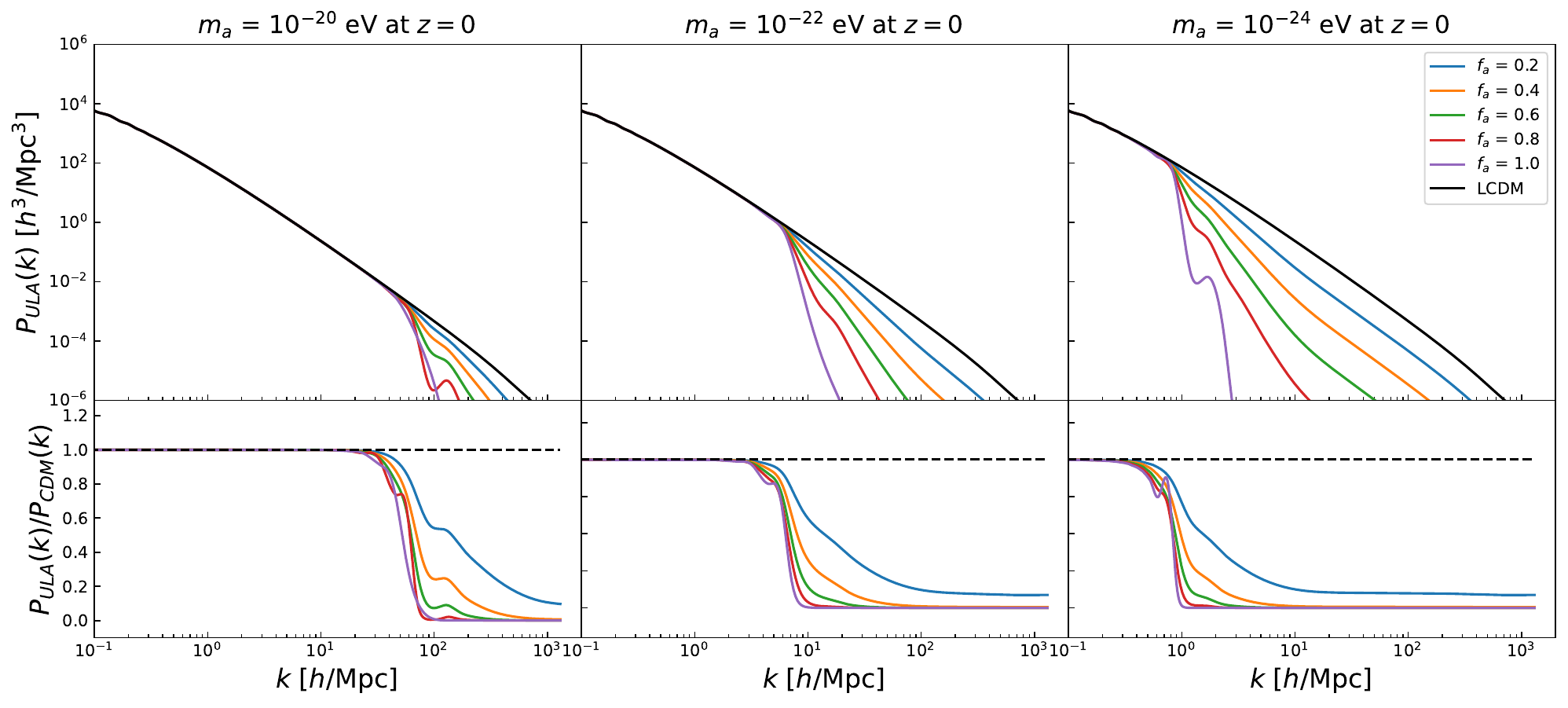}
    \caption{The linear matter power spectra for different ULA masses $m_{\rm a}$ and ULA fractions $f_{\rm a}$. We can see that the amplitude and scale of the suppression are dependent on $f_{\text{a}}$ and  $m_{\text{a}}$, respectively. The ratio of the ULA and CDM matter power spectra are also shown in the lower panels.}
    \label{fig:matter-ps}
\end{figure*}

The ULAs are scalar bosons, and can be described by a non-relativistic field $\phi$, and the equation of motion for the background axion field is
\begin{equation}
    \ddot{\phi}_{0}+2 \mathcal{H} \dot{\phi}_{0}+m_{\text{a}}^{2} a^{2} \phi_{0}=0,
\label{eq:axion_1}
\end{equation}
where the conformal Hubble parameter $\mathcal{H} = \dot{a}/a = aH$, and $\phi_{0}(\tau)$ is the homogeneous value of the scalar field as a function of the conformal time $\tau$. Then the corresponding density and pressure of the background axion field are given by \citep{axionCAMB},
\begin{equation}
    \rho_{\text{a}}=\frac{1}{2 a^{2}} \dot{\phi}_{0}^{2}+\frac{m_{\text{a}}^{2}}{2} \phi_{0}^{2},
\label{eq:axion_2}
\end{equation}
\begin{equation}
    P_{\text{a}}=\frac{1}{2 a^{2}} \dot{\phi}_{0}^{2}-\frac{m_{\text{a}}^{2}}{2} \phi_{0}^{2}.
\label{eq:axion_3}
\end{equation}

At early times ($H \gg m_{\text{a}}$), the axion field is in the slowly rolling phase with $\phi_{0} \thickapprox 0$, and then the equation of state of axion is $w_{\text{a}} \equiv P_{\text{a}}/\rho_{\text{a}} \thickapprox -1$. So the axion field behaves like a dark energy component at early times. At late times ($H \ll m_{\text{a}}$), the ULA equation of state oscillates around zero, and the energy density of the field obeys $\rho_{\text{a}} \propto a^{-3}$. Then the ULA acts like dark matter in this phase. Therefore, the scale factor when the transition from dark energy to dark matter happens is critical, and we denote it as $a_{\text{osc}}$, which satisfies $H(a_{\text{osc}}) \thickapprox m_{\text{a}}$, i.e. the time of this transition is dependent on the ULA mass. For $m_{\text{a}} \lesssim 10^{-28}$ eV, we have $a_{\text{osc}} > a_{\text{eq}}$. Then the axion field does not behave as dark matter at matter-radiation equality, and cannot constitute the entirety of the dark matter. For $m_{\text{a}} \gtrsim 10^{-28}$ eV, the ULA can act as dark matter, and we mainly focus on these dark-matter-like axions in this work.

One of the main properties of the ULAs is that they suppress small-scale gravitational clustering due to their macroscopic de Broglie wavelengths. We can define a characteristic Jeans scale $k_{\text{J}}$ in Fourier space, above which axions can not cluster. For the linear matter power spectrum with dark-matter-like ULAs, the suppression is frozen in at matter-radiation equality, and we have \citep{Hu-2000},
\begin{equation}
    k_{\text{J,eq}} \thickapprox 9 \left( \frac{m_{\text{a}}}{10^{-22}\text{ eV}} \right)^{1/2} \text{ Mpc}^{-1}.
\label{eq:axion_4}
\end{equation}
On the other hand, the suppression on small scales not only depends on the axion mass, but also is related to the ULA fraction $f_{\text{a}} = \Omega_{\text{a}}/\Omega_{\text{d}}$. When $\Omega_{\text{a}} \to 0$, it shall recover the pure CDM result. In this work, the linear matter power spectrum which includes axion effects is computed using the Boltzmann code {\tt axionCAMB} \citep{axionCAMB}, which is a modified version of the code {\tt CAMB} \citep{CAMB}.

In Figure~\ref{fig:matter-ps} we show the linear matter power spectra for different axion masses $m_{\text{a}}$ and $f_{\text{a}}$, obtained by {\tt axionCAMB}. We note that, the amplitude of the suppression is mainly dependent on the ULA fraction $f_{\text{a}}$, i.e. the higher this fraction, the stronger the suppression is. On the other hand,  the suppression scale is dependent on the axion mass $m_{\text{a}}$, that heavier axions will only significantly suppress  matter fluctuations at smaller physical scales. Since the mass range of $10^{-33} \lesssim m_{\text{a}} \lesssim 10^{-24}$ eV has been ruled out by the CMB observation \citep{Hlozek-2018}, especially for the low-mass range $m_{\text{a}} \lesssim 10^{-27}$ eV, we will mainly focus on the mass range from $10^{-26}$ eV to $10^{-18}$ eV in our work.

\subsection{Halo Model and Non-linear Power Spectrum}\label{subsec:Halo-Model}

Since we consider a relative high ULA masses, the suppression effect from the ULAs will mainly affect the small scales. Therefore,  we need to accurately calculate the non-linear matter power spectrum including the ULA effect. In our work, we modify the {\tt HMCode} 2020 vesion \citep{Mead-2020}, and adapt it for the ULA cosmology. 

The key concept of the halo model is assuming all matter is associated with virialized dark matter halos, and hence the statistical properties of the cosmic structure (especially, in non-linear region) can be described through modeling the spatial distribution of these halos and the distribution of dark matter within them \citep{Cooray-2002}. To model the halo distribution, first we need to calculate the halo mass function $n(M)$. Here we adopt the Press-Schechter (PS) approach \citep{Press-Schechter} :
\begin{equation}
    n(M)dM = \frac{\bar{\rho}}{M} f(\nu) d\nu,
    \label{eq:dndm}
\end{equation}
where $\bar{\rho}$ is the mean comoving matter density, $\nu(M) = \delta_{\text{c}}/\sigma(M)$ denotes peak height, related with the variance of the linear matter overdensity field $\sigma(M)$ and the critical overdensity barrier for collapse $\delta_{\text{c}}$. Then the function $f(\nu)$ is written as \citep{Sheth-Tormen-1999, Sheth-Tormen-2001, Cooray-2002}:
\begin{equation}
    f(\nu) = A \left[ 1 + (q\nu^2)^{-p} \right]e^{-q\nu^2/2},
    \label{eq:fv}
\end{equation}
with $p = 0.3$, $q = 0.707$, $A = 0.2161$. The variance of linear power spectrum can be calculated by
\begin{equation}
    \sigma^2(R) = \int_{0}^{\infty} 4\pi \left(\frac{k}{2\pi}\right)^3P^{\text{lin}}(k)W^2(kR)d{\rm \,ln}k,
    \label{eq:sigma-R}
\end{equation}
where $W(x) = (3/x^3)\left(\sin x - x\cos x\right)$ is the Fourier transform of a spherical top-hat filter window function. We can rewrite $\sigma(R)$ in terms of mass by relating the comoving scale $R$ with the the mass enclosed within this scale as $M = (4/3)\pi R^3 \bar{\rho}$.

In the case of $\Lambda$CDM model, the critical overdensity barrier at $z = 0$ is found to be $\delta_{\text{c}}^{0} \approx 1.686$. In the formalism of peak height, we can absorb the linear growth factor $D(z)$ of the linear power spectrum into the overdensity barrier, and obtain a redshift dependent barrier, which is given by
\begin{equation}
    \delta_{\text{c}}^{\text{CDM}}(z) = \frac{\delta_{\text{c}}^{0}}{D_{\rm CDM}(z)}.
    \label{eq:delta_c_z}
\end{equation}

For the ULA case, the suppression on the matter fluctuations via its quantum pressure is expected to cause a scale-dependent perturbation growth, which can lead to a mass-dependent overdensity barrier as
\begin{equation}
    \delta_{\text{c}}^{\text{ULA}}(M,z) = \frac{D_{\text{CDM}}(z)}{D_{\text{ULA}}(M,z)} \delta_{\text{c}}^{\text{CDM}}(z),
    \label{eq:delta_c}
\end{equation}
where $D_{\text{ULA}}(M,z)$ is the mass-dependent growth factor for the ULA. Following \cite{Marsh-2014}, the relative amount of growth between ULA and CDM can be calculated as
\begin{equation}
\begin{aligned}
    &\mathcal{G}(k,z) \equiv \frac{D_{\text{CDM}}(z)}{D_{\text{ULA}}(M,z)} \\
    & = \frac{\delta_{\text{CDM}}(k,z) \delta_{\text{CDM}}(k_0,z_{\text{h}}) \delta_{\text{ULA}}(k,z_{\text{h}}) \delta_{\text{ULA}}(k_0,z)}{\delta_{\text{CDM}}(k,z_{\text{h}}) \delta_{\text{CDM}}(k_0,z) \delta_{\text{ULA}}(k,z) \delta_{\text{ULA}}(k_0,z_{\text{h}})},
    \label{eq:G_k_z}
\end{aligned}
\end{equation}
where $k_{0} = 0.002 h$ Mpc$^{-1}$ is the pivot scale, and $z_{\text{h}}$ should be chosen to be the redshift that the transfer function of $\Lambda$CDM has frozen in. In \cite{Marsh-2014}, they suggested $z_{\text{h}} = 300$ will work well for both CDM and ULA cases. $\delta(k,z)$ is the matter overdensity in Fourier space, and its numerical value for the ULA case can be  obtained by {\tt axionCAMB} \citep{axionCAMB}.

\begin{figure}
    \centering
    \includegraphics[scale = 0.55]{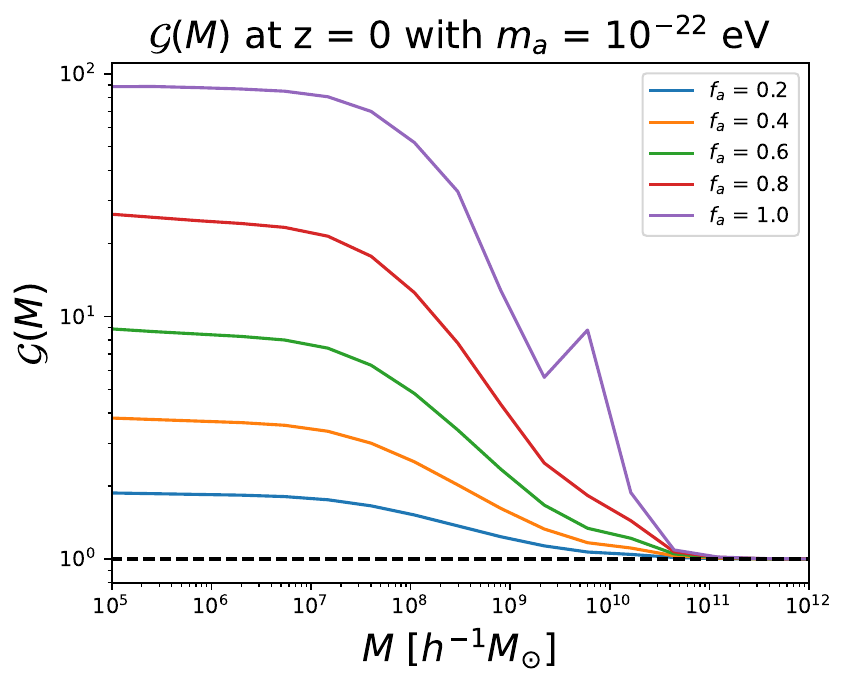}
    \caption{The mass-dependent factor $\mathcal{G}(M)$ for $m_{\text{a}} = 10^{-22}$ eV and different ULA fraction $f_{\text{a}}$ at $z = 0$.}
    \label{fig:G_k}
\end{figure}

\begin{figure}
    \centering
    \includegraphics[scale = 0.56]{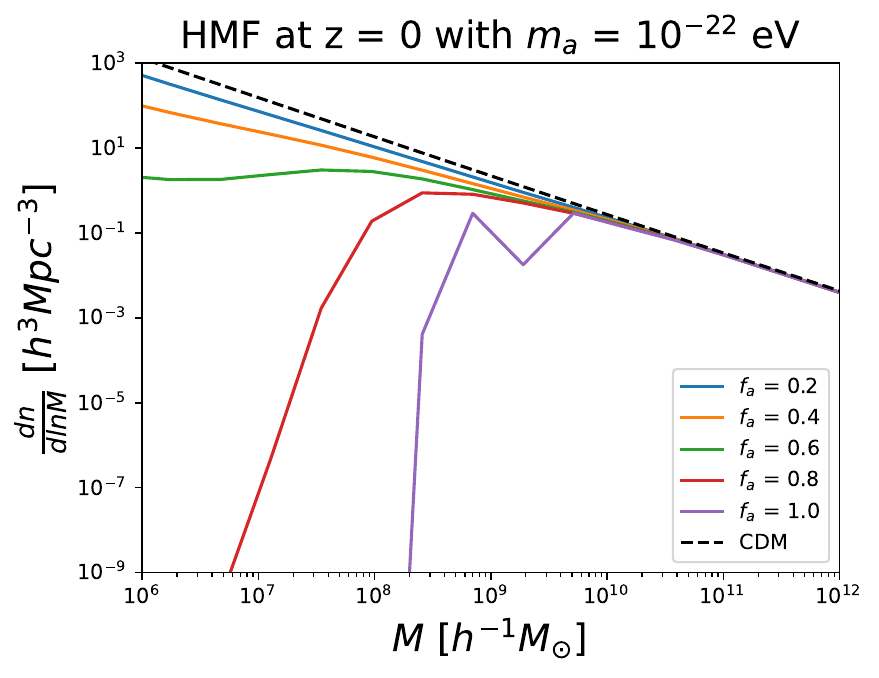}
    \caption{The halo mass function for $m_{\text{a}} = 10^{-22}$ eV and different ULA fraction $f_{\text{a}}$ at $z = 0$. The black dash line shows the $\Lambda$CDM case for comparison.}
    \label{fig:dndm}
\end{figure}

In Figure~\ref{fig:G_k} we show the mass-dependent factor $\mathcal{G}(M)$ for $m_{\text{a}} = 10^{-22}$ eV with different $f_{\text{a}}$ from 0.2 to 1.0 at $z = 0$. Obviously, $\mathcal{G}(M)$ increases at low halo masses or small scales given an ULA mass, and higher $f_{\rm a}$ leads to higher $\mathcal{G}(M)$, since higher $f_{\rm a}$ has stronger suppression at small scales. 
In Figure~\ref{fig:dndm}, we plot the halo mass function for $m_{\text{a}} = 10^{-22}$ eV with various $f_{\text{a}}$ at $z = 0$, and the $\Lambda$CDM case is also shown in black dashed line for comparison.We can see that, the suppression of the halo mass function begins at the same mass scale when the ULA mass is fixed, and the amplitude of suppression increases with the raising of the ULA fraction.

We notice that there is a spike-like oscillatory feature in $\mathcal{G}(M)$ and halo mass function for $f_{\text{a}} = 1$ as shown in Figure~\ref{fig:G_k} and Figure~\ref{fig:dndm}. The origin of this shape is due to that the overdensity will become vanishingly small at low-mass scales (or equivalently, at high-k scales), when the matter field is completely dominated by the ULA (i.e. $f_{\text{a}} = 1$). Then in Eq.~(\ref{eq:G_k_z}), a numerical instability problem of dividing zero by zero will appear. As proposed by \cite{Marsh-2014}, the combination of this numerical issue and the BAO distortions effect can lead to the spikey feature for the $f_{\text{a}} = 1$ case. As we show later, this effect would not affect our result.

\begin{figure}
    \centering
    \includegraphics[scale = 0.52]{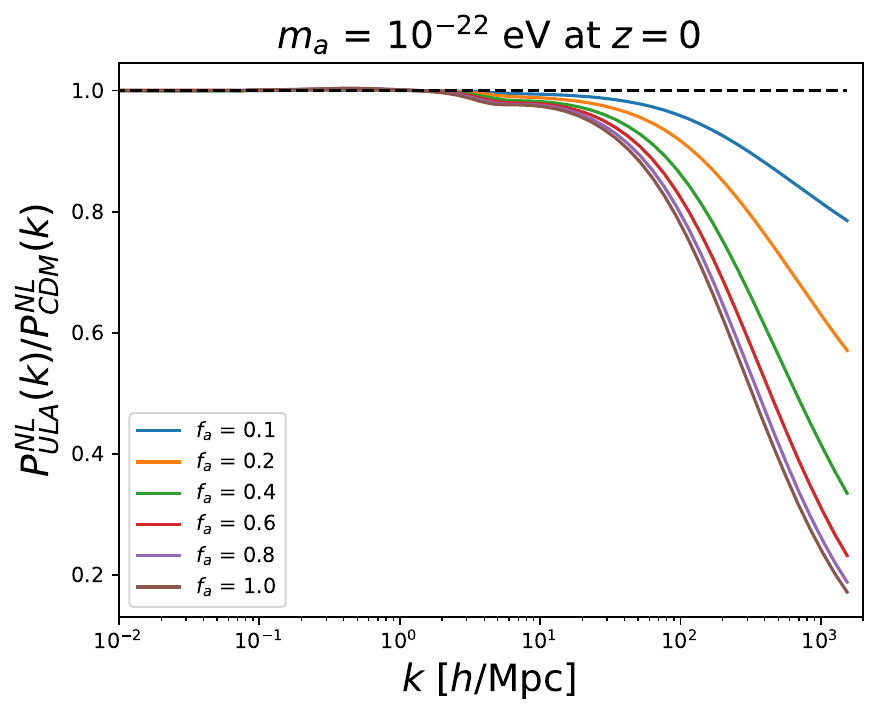}
    \caption{The ratio of the ULA and CDM non-linear power spectra for $m_{\rm a}=10^{-22}$ eV at $z=0$, considering different ULA  fraction $f_{\rm a}$.}
    \label{fig:nonlinearPS}
\end{figure}

In order to calculate the non-linear power spectrum including the ULAs, we use the pure-Python implementation of the HMCode-2020 \citep{Mead-2020}, i.e. {\tt HMCode-python}\footnote{https://github.com/alexander-mead/HMcode-python}, by inputting the linear power spectrum and halo mass function with the ULAs obtained above into the code. In Figure~\ref{fig:nonlinearPS}, we show the ratio of the ULA and CDM non-linear power spectra for $m_{\rm a}=10^{-22}$ eV at $z=0$ with different values of $f_{\rm a}$.

Comparing to other models, for example, \cite{Vogt-2023} presents an alternative model to describe the non-linear clustering behavior of mixed ULA cosmology (i.e. ULA not contribute to all dark matter). In their model, they assume a ULA halo form around a cold dark matter halo with mass $M_{\text{ax}} = (\Omega_{\text{ax}}/\Omega_{\text{c}})M_{\text{c}}$ (which called the ULA halo mass relation), and the density profile of ULA halo is given by a soliton core and an NFW profile in the outer regions, which makes major impact to the one halo term of the halo model calculation. Hence this model would be quite useful when trying to explore very small scales. Their result is confirmed by \cite{Lague-2023}, who run the first mixed ULA cosmological simulation to investigated the non-linear power spectrum and the halo density profile for mixed dark matter scenario. We find that the suppression on the non-linear matter power spectrum in our model is similar with that given by \cite{Vogt-2023}. Since the shapes of the curves only have slightly difference, we believe the impact on the constraint result is negligible, and we will adopt our current model in the analysis.

\subsection{Baryonic Feedback}

Besides the non-linear clustering of dark matter when exploring the small scales, we also consider the effect of baryonic feedback. The baryon, which contributes about one sixth of the total matter, has significant impact at the non-linear scales. The impact of baryon, known as baryonic feedback, has two main processes. First, the contraction of radiative cooling gas will alter the dark matter distribution via gravitational force, and then causes the change of the distribution of all matter \citep{Duffy-2010}; Second, the supernova explosion or the active-galactic nuclei (AGN) will release a huge amount of energy and matter, which can push gas to the outskirt of dark matter halos \citep{Schaye-2010, Chisari-2018, vanDaalen-2020}. The detail of those mechanisms still are not well-understood, but it has been well-demonstrated by high-resolution hydrodynamic simulations. So we can phenomenologically fitting the parameterized baryonic feedback model to match the result from hydrodynamic simulations, such as COSMO-OWLS \citep{COSMO-OWLS} and BAHAMAS \citep{BAHAMAS}, and then these models can be used to study the impact of the baryonic feedback effect.

In \cite{Mead-2020-no-Baryon}, they introduced a six-parameter model to describe the baryonic feedback. Each of these parameters has clear physical motivation, including the halo concentration parameter $B$, the effective halo stellar mass fraction $f_{*}$ related to the power spectra of stellar matter, the halo mass threshold $M_{\text{b}}$ denoting haloes with $M < M_{\text{b}}$ that have lost more than half of their gas, and the redshift evolution of these three parameters modeled by $B_{z}$, $f_{*,z}$ and $M_{\text{b},z}$. However, further investigation, as shown in \cite{Mead-2020}, pointes out that there is some degeneracies between those six parameters, and all of them can be linearly fitted as a functions of $\text{log}_{10}(T_{\text{AGN}}/\text{K})$. $T_{\text{AGN}}$ is called as AGN temperature, which describes the strength of feedback. But we should note that $T_{\text{AGN}}$ is actually not a real physical observable in real observations.

Note that the baryonic feedback model presented by \cite{Mead-2020} is calibrated by pure CDM simulation. In order to describe the impact of baryonic feedback in our case, an ULA-base simulation is actually needed. For the ULA cosmology, \cite{Mocz-2019} pointed out that the feature of dark matter distribution may not be significantly affected by baryonic feedback. \cite{Veltmaat-2020} ran a zoom-in simulation, and found that the velocity dispersion of dark matter in the center of halo will increase, and the mass-radius relation can be altered when including baryonic feedback, resulting in a more massive and compact core of dark matter halo, which can have significant effects at very small scales. However, since there are no large-scale cosmological simulation in the ULA scenario with the present of baryonic feedback, we assume the baryonic feedback model presented by \cite{Mead-2020} is still available in our case.

In our work, we adopt this single-parameter model to include the baryonic feedback effect. \cite{BAHAMAS} found that the AGN temperature with $\text{log}_{10}(T_{\text{AGN}}/\text{K}) = 7.8$ can reproduce a simulation result which has good agreement with  observed galaxy stellar mass function and hot gas mass fractions. So we will set this value as the fiducial value, and adopt a prior range of $7.4 < \text{log}_{10}(T_{\text{AGN}}/\text{K}) < 8.3$, which is recommended by \cite{Mead-2020}.

\section{Mock Data}\label{sec:Mock-data}

\subsection{Galaxy Photometric Redshift Distribution}\label{subsec:z-dist}

The first step for generating the {\it CSST} mock photometric data is to estimate the galaxy redshift distribution. Here we adopt a catalog based on the {\it CSST} instrumental design and COSMOS catalog \citep{COSMOS1, COSMOS2}, which is suggested by \citet{Cao2018}. This catalog contains about 220,000 sources in a 1.7 deg$^2$ field, by assuming a galaxy detection of $\le 25.2$, which is a similar magnitude limit as the {\it CSST} photometric survey. As shown in \citet{Cao2018}, the galaxy redshift distribution of the {\it CSST} photometric survey has a peak at $z=0.6$, and the redshift range can cover up to $z\sim4$.

\begin{figure}
    \includegraphics[scale = 0.46]{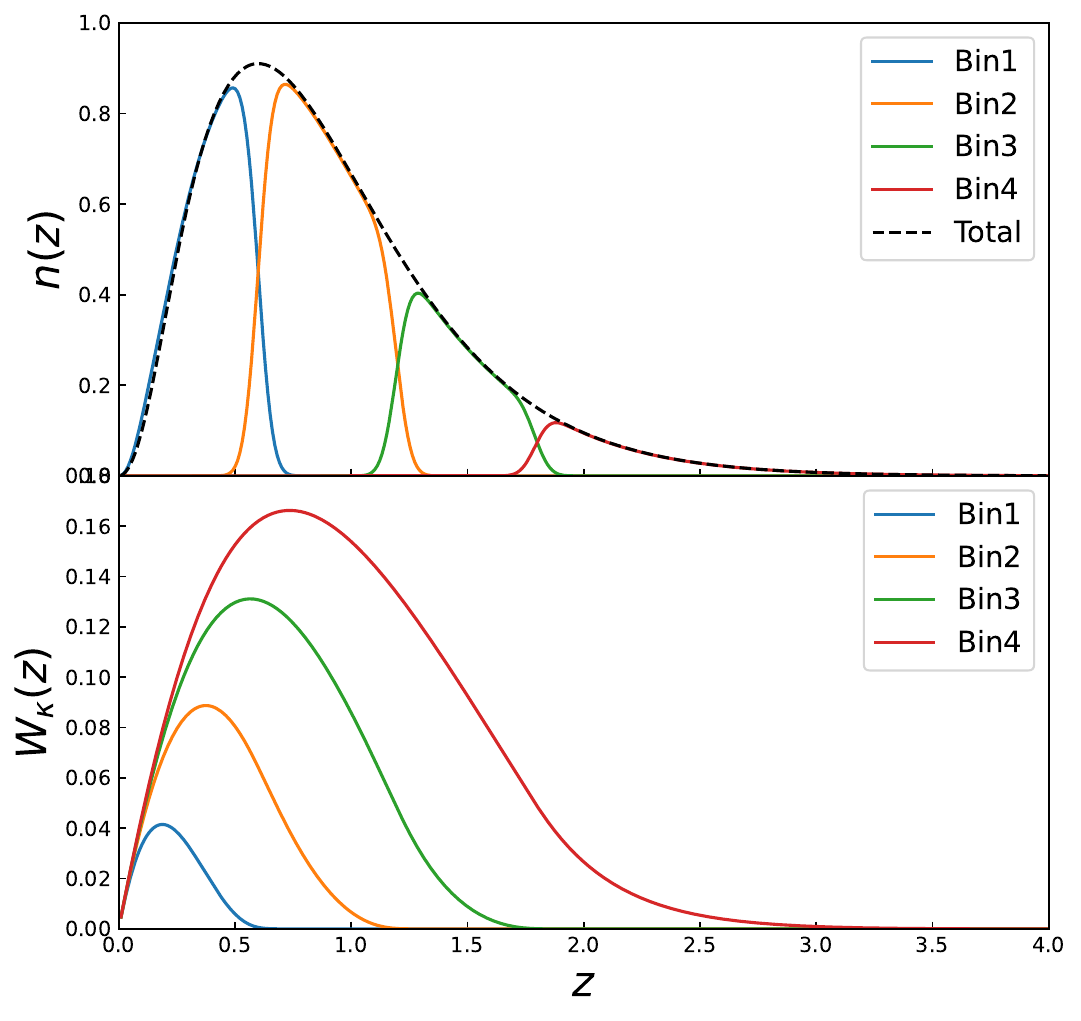}
    \caption{{\it Upper panel}: galaxy redshift distribution of the CSST photometric surveys. The black dotted curve shows the total redshift distribution $n(z)$, and the solid blue, orange, green and red curves show the $n_i(z)$ for the four tomographic bins, with $\Delta_z = 0$ and $\sigma_z = 0.05$. {\it Lower panel}: the weighting kernels of the four tomographic bins in the CSST cosmic shear survey.}
    \label{fig:z-dist}
\end{figure}

For simplicity, here we use an analytical smooth function to represent this redshift distribution, which takes the form as
\begin{equation}
    n(z) \propto z^2e^{-z/z^{*}}.
    \label{eq:z-dist-2}
\end{equation}
In our case, $z^* = z_{\rm peak}/2 = 0.3$. In the upper panel of Figure~\ref{fig:z-dist}, the normalized $n(z)$ with $\int n(z){\rm d}z=1$ is shown in black dashed curve. And we can find that our redshift distribution can well match the distribution given in \citet{Cao2018}.

In order to extract more information from the photometric data, as shown by the solid curves, we divide our redshift distribution into four tomographic bins, by assuming the photo-$z$ bias $\Delta_z=0$, and photo-$z$ scatter $\sigma_z=0.05$, which is suggested by \citet{Gong-CSST-2019}. Then we can measure the auto and cross galaxy clustering or shear power spectra for these bins. In principle, more tomographic bins can basically further improve the constraint result, for example, \citet{Gong-CSST-2019} found that six photo-$z$ bins can improve the constraints by a factor of $\sim1.5$. So in the real data analysis we can use more bins, but in the current stage, four tomographic bins are sufficient for this work.

\subsection{Shear and Galaxy Angular Power Spectra}\label{subsec:Observables-and-Model}

\begin{figure*}
	\includegraphics[scale = 0.52]{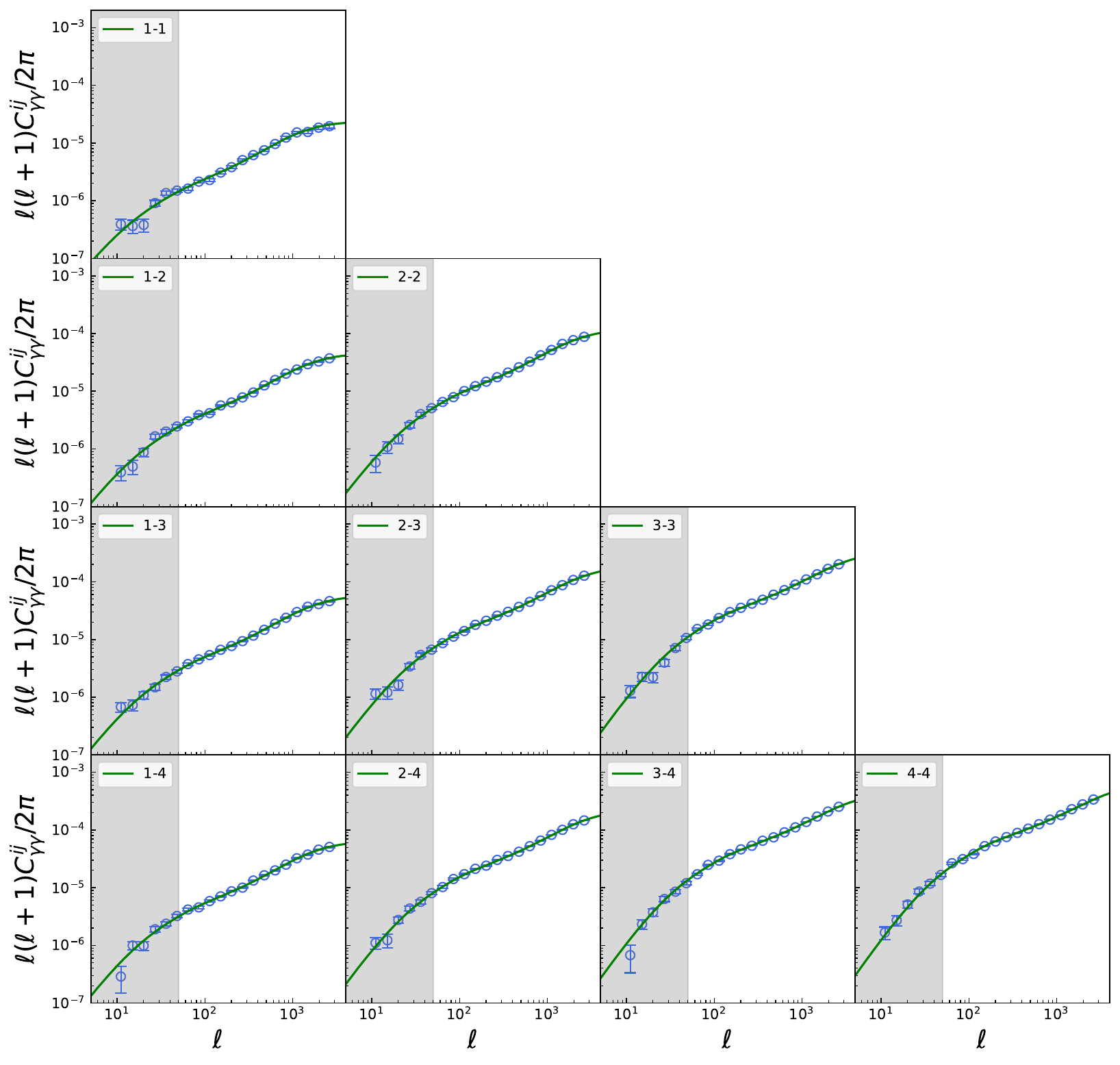}
    \caption{The mock auto and cross signal power spectra for the four tomographic bins in the CSST cosmic shear survey. The green curves and blue data points denote the theoretical models and mock data, respectively. The gray regions exclude the scales at $\ell<50$ where the Limber approximation may not be available.}
    \label{fig:shear_power_spectrum}
\end{figure*}

\begin{figure*}
	\includegraphics[scale = 0.52]{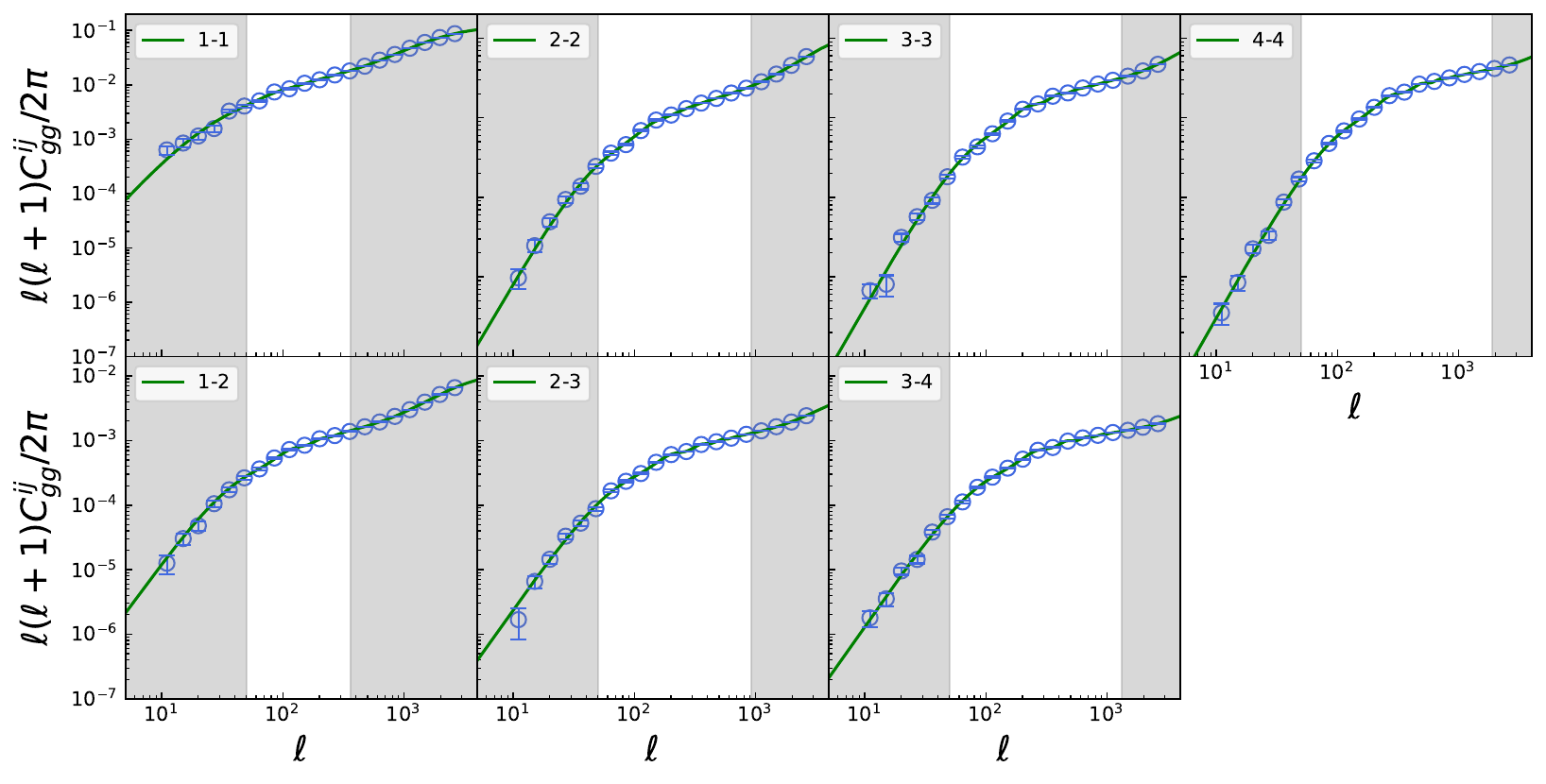}
    \caption{The mock CSST angular galaxy power spectra of the four tomographic bins. The green solid curves show the results of the fiducial theoretical model, and blue data points are the mock data. Since the cross power spectra between nonadjacent bins are quite small, we only consider three cross power spectra between adjacent bins with significant amplitudes. The gray regions show the scales for excluding the Limber approximation and non-linear effects.}
    \label{fig:gg_power_spectrum}
\end{figure*}

\begin{figure*}
	\includegraphics[scale = 0.52]{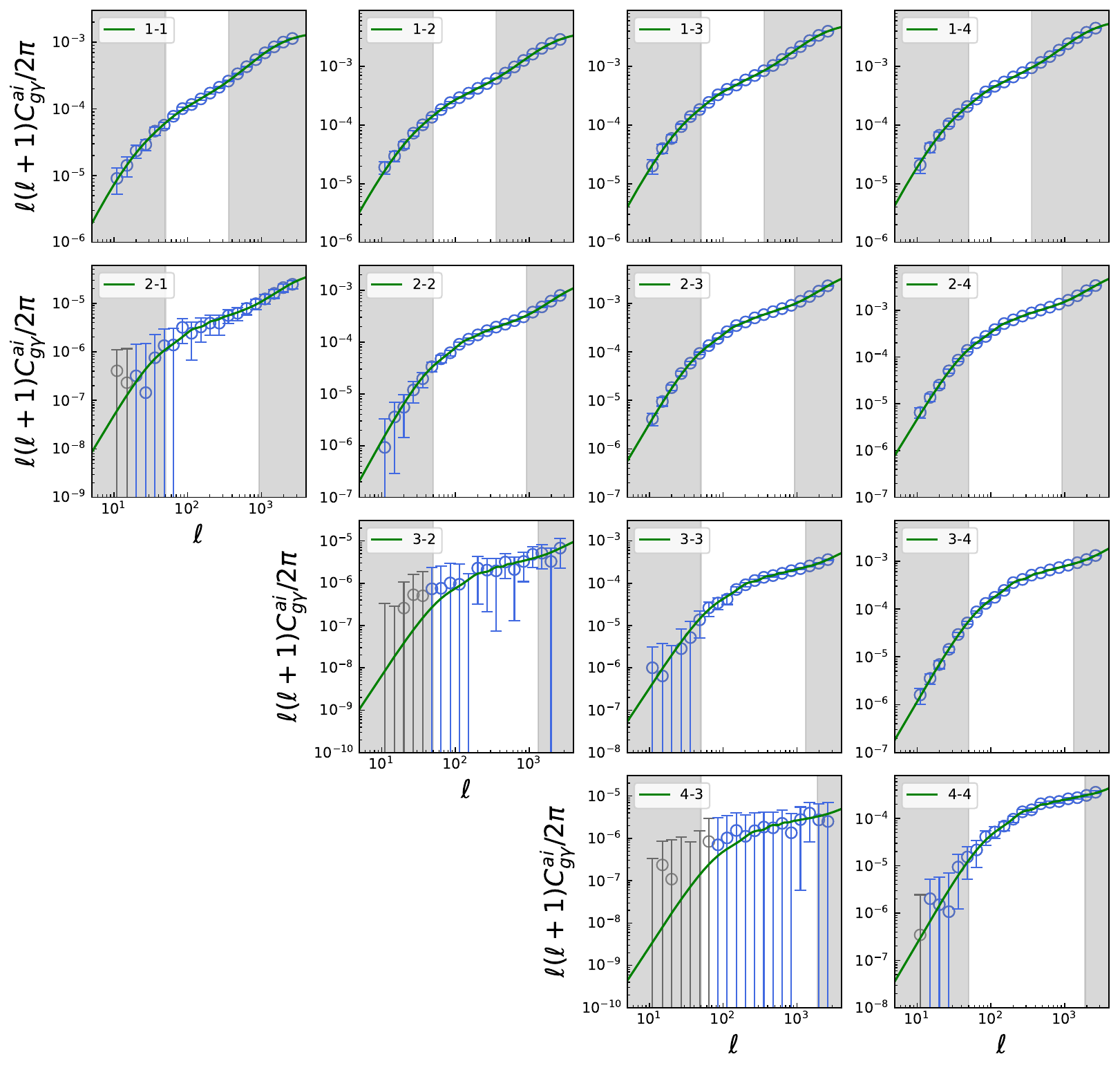}
    \caption{The mock CSST galaxy-galaxy lensing power spectra of the four tomographic bins. The green curves and blue data points denote the theoretical models and mock data, respectively. We discard the cross power spectra with low amplitudes, and only consider the data points with $S/N \ge 1$ (blue data points). The gray regions denote the excluded scales, which are mainly restricted by the galaxy clustering survey.}
    \label{fig:cross_power_spectrum}
\end{figure*}

We model the shear and galaxy angular power spectra based on a given set of cosmological and systematical parameters. Assuming Limber approximation \citep{Limber}, the general angular power spectra for shear and galaxy surveys can be written as
\begin{equation}
    C_{\text{uv}}^{ij}(\ell) = \frac{1}{c} \int dz H(z)D_{\text{A}}^{-2}(z)W_{\text{u}}^{i}(z)W_{\text{v}}^{j}(z)P_{\text{m}}\left(\frac{\ell + 1/2}{D_{\text{A}}(z)}, z\right),
    \label{eq:C_ell-1}
\end{equation}
where u, v $\in \{ \text{g, }\kappa \text{,}\ I \}$ denote different tracers, and g, $\kappa$ and $I$ stand for galaxy clustering, cosmic shear and  intrinsic alignment, respectively. $c$ is the speed of light, and $D_{\text{A}}$ is the comoving angular diameter distance. 
The weighting function of galaxy clustering is given by
\begin{equation}
    W_{\text{g}}^i(z) = b_{i} n_i(z),
\label{eq:gg_kernel}
\end{equation}
where $n_{i}(z)$ is the normalized redshift distribution of the $i$th bin (as we shown in the upper panel of Fig. \ref{fig:z-dist}), and $b_{i}$ is the linear galaxy clustering bias, which connects the clustering of galaxy with the underlying matter density field. We assume it varies with redshift following: $b(z) = 1 + 0.84z$ \citep{Weinberg-2004}. In our model, we assume the galaxy bias is a constant in each tomographic bin, with value of $b_{i} = 1 + 0.84z_{\text{cen}}^{i}$, where $z_{\text{cen}}^{i}$ is the central redshift of a redshift bin.

\begin{table}
    \centering
        \caption{Free parameters considered in the constraint process. The first column shows the names of our 28 free parameters. The second and third columns show the fiducial values and the prior ranges of the parameters.}

    \begin{tabular}{lcccccr}
       \hline
       \text{Parameter} & \text{Fiducial Value} & \text{Prior}  \\
        
       \hline
       \text{Cosmological Parameter} \\
       \hline
       $\Omega_{\text{m}}$ & 0.32 & flat (0, 0.6)  \\
       $\Omega_{\text{b}}$ & 0.048 & flat (0.01, 0.1) \\
       h & 0.6774 & flat (0.4, 1.0)  \\
       $n_{\text{s}}$ & 0.96 & flat (0.8, 1.2)  \\
       $w$ & -1 & flat(-2, 0)  \\
       $\Sigma m_{\nu}$ & 0.06 eV & flat(0, 2) eV \\
       
       $\sigma_8$ & 0.8 & flat (0.4, 1.2)  \\
       \hline
       \text{Baryonic Effect} \\
       \hline
       $\text{log}_{10}(T_{\text{AGN}}/\text{K})$ & 7.8 & flat (7.4, 8.3)   \\
       \hline
       \text{ULA Parameter} \\
       \hline
       $f_{\text{a}}$ & - & flat (0, 1.0) \\
       $\text{log}_{10}(m_{\text{a}}/\text{eV})$ & - & flat (-26, -18) \\
       \hline
       \text{Intrinsic Alignment} \\
       \hline
       $A_{\text{IA}}$ & 1 & flat (-5, 5)  \\
       $\eta_{\text{IA}}$ & 0 & flat (-5, 5) \\
       \hline
       \text{Galaxy Bias} \\
       \hline
       $b^i$ & (1.252,1.756,2.26,3.436) & flat (0, 5)  \\
       \hline
       \text{Photo-z Bias} \\
       \hline
       $\Delta z^i$ & (0,0,0,0) & flat (-0.1, 0.1)  \\
       $\sigma_z^i/\sigma_{z,\text{fid}}^i$ ($\sigma_{z,\text{fid}}=0.05$) & (1,1,1,1) & flat (0.5, 1.5)\\
       \hline
       \text{Shear Calibration} \\
       \hline
       $m_i$ & (0,0,0,0) & flat (-0.1, 0.1) \\
       \hline
    \end{tabular}
    \label{tab:params-1}
\end{table}

For cosmic shear measurement, the weighting function can be written as \citep{Hu&Jain2004}
\begin{equation}
    W_{\kappa}^i(z) = \frac{3\Omega_{\rm m} H_0^2}{2H(z)c}\frac{D_{\text{A}}(z)}{a} \int_z^{\infty} dz^{\prime} n_i(z^{\prime}) \frac{D_{\text{A}}(z, z^{\prime})}{D_{\text{A}}(z^{\prime})}.
\label{eq:shear_kernel}
\end{equation}
In the lower panel of Fig. \ref{fig:z-dist} , we show the lensing kernel for the four tomographic bins. We can see that the distribution of the shear weighting function is basically wider than the corresponding galaxy redshift distribution, especially for the tomographic bins at higher redshifts. 

The intrinsic alignment effect arises from the local gravitational potential can influence the cosmic shear measurement, which will be a major systematic of weak lensing survey. We model it as \citep{Hildebrandt-2017}
\begin{equation}
    W_{\text{I}}^i(z) = A_{\text{IA}}C_1 \rho_{\text{c}} \frac{\Omega_{\text{m}}}{D(z)} n_i(z) \left( \frac{1+z}{1+z_0} \right)^{\eta_{\text{IA}}} \left( \frac{L_i}{L_0} \right)^{\beta_{\text{IA}}},
\label{eq:IA_kernel}
\end{equation}
where $C_1 = 5 \times 10^{-14} h^{-2}$ $\text{M}_{\odot}^{-1}\text{Mpc}^3$ is a constant, $\rho_{\text{c}}$ is the critical density of present day, $D(z)$ is the linear growth factor which is normalized to unity at $z = 0$. $A_{\text{IA}}$, $\eta_{\text{IA}}$ and $\beta_{\text{IA}}$ are free parameters, and $z_0$ and $L_0$ are pivot redshift and luminosity. For simplicity, we ignore the luminosity dependence here, by fixing $\beta_\text{{IA}} = 0$. And we set $z_0 = 0.6$, and take of $A_{\text{IA}}=1$ and $\eta_{\text{IA}}=0$ as the fiducial values.

Then the angular power spectra of galaxy, shear, and galaxy-galaxy lensing measurements, considering systematical effects, can be calculated as \citep{Abbott-2018}
\begin{equation}
    \Tilde{C}_{\text{gg}}^{ab} = C_{\text{gg}}^{ab} + \delta_{ab}\frac{\sigma_{g}^2}{\Bar{n}_a} + N_{\text{sys}}^{\text{g}},
\label{eq:C_gg}
\end{equation}

\begin{equation}
\begin{split}
    \Tilde{C}_{\gamma \gamma}^{ij} = (1+m_i)(1+m_j) \left[ C_{\kappa\kappa}^{ij}(\ell) + C_{\text{II}}^{ij}(\ell) +
    C_{\kappa\text{I}}^{ij}(\ell) + C_{\kappa\text{I}}^{ji}(\ell) \right]\\
    + \delta_{ij}\frac{\sigma_{\gamma}^2}{\Bar{n}_i}+N_{\text{sys}}^{\gamma},
    \label{eq:shear}
\end{split}
\end{equation}

\begin{equation}
    C_{\text{g}\gamma}^{ai}(\ell) = C_{\text{g}\kappa}^{ai}(\ell) + C_{\text{gI}}^{ai}(\ell).
    \label{eq:ggl}
\end{equation}
Here we use $a$ and $b$ to denote the redshift bins of galaxy samples, and $i$ and $j$ for the cosmic shear samples, which will be more distinct for calculating galaxy-galaxy lensing signal using equation (\ref{eq:ggl}). The second to last terms in equation (\ref{eq:C_gg}) and equation (\ref{eq:shear}) represent shot noise, where $\delta _{ab}$ and $\delta _{ij}$ are Kronecker delta. For galaxy clustering survey, we have $\sigma_{g} = 1$, and for weak lensing survey, we assume $\sigma_{\gamma} = 0.2$, which is due to the intrinsic shape of galaxies and measurement errors. The last term in equation (\ref{eq:C_gg}) and equation (\ref{eq:shear}) are additive systematic errors. For galaxy clustering, it comes from spatially varying dust extinction, instrumentation effects, etc \citep{Tegmark-2002, Zhan06}. For weak lensing survey, it arises from PSF, instrumentation effects, and so on \citep{Guzik-2005, Jain-2006}. In our work, we set $N_{\text{sys}}^{\text{g}} = 10^{-8}$ and $N_{\text{sys}}^{\gamma} = 10^{-9}$ for the CSST surveys, which are assumed to be independent on redshift or scale \citep{Gong-CSST-2019}.

We use 20 log-spaced bins between $10 \leqslant \ell \leqslant 3000$ to generate our mock data. We discard the data at scales $\ell < 50$ to avoid the break down of Limber approximation on large physical scales \citep{Fang-2020}. For weak lensing, we use the power spectra up to $\ell = 3000$\footnote{If using a larger scale cut-off, e.g. $\ell\lesssim2000$, we find that the constraints on the ordinary cosmological parameters, such as $\Omega_{\rm m}$ and w, will become a bit looser ($\sim$10\%) compared to that for $\ell<3000$. However, the impact on the constraints of $m_{\rm a}$, $f_{\rm a}$ and $m_{\nu}$ is significant ($\sim$20\%-60\% looser), since they are sensitive to the feature at small scales.}. And for galaxy clustering and galaxy-galaxy lensing measurements, to avoid the uncertainty of the non-linear effect at small scales, we  set a maximum wavenumber scale $k_{\text{max}} = 0.3$ $h$ Mpc$^{-1}$ \citep{LSST-Science-Requirements, Wenzl-2022}, and relate it to $\ell$-space via $\ell = k\chi(z)$.

In Fig. \ref{fig:shear_power_spectrum}, Fig. \ref{fig:gg_power_spectrum} and Fig. \ref{fig:cross_power_spectrum}, we show the mock data of the CSST cosmic shear, angular galaxy clustering and galaxy-galaxy lensing power spectra, respectively. The gray regions show the scales for excluding the break down of Limber approximation (small $\ell$ regions) and non-linear effects (high $\ell$ regions). For cosmic shear survey, we present all ten auto and cross signal power spectra for the four tomographic bins in Fig \ref{fig:shear_power_spectrum}. But for angular galaxy clustering case, we notice that only the two adjacent bins have the cross power spectrum, since for the given the redshift scatter we assume, there is no overlapping for galaxy redshift distribution in other cases. So we totally have seven auto and cross galaxy angular power spectra as shown in Fig. \ref{fig:gg_power_spectrum}. For the galaxy-galaxy lensing case, $C_{\text{g}\gamma}^{ai}$ only has significant signal when $a \leqslant i$ as indicated in Fig.~\ref{fig:z-dist}. So there is a strong correlation between the background shear and the foreground galaxy clustering; on the contrary, the low redshift shear signal has less correlations with the background matter distribution. After excluding the data point with $S/N < 1$, only 13 galaxy-galaxy lensing power spectra for the four tomographic bins are obtained.

\subsection{Covariance Matrix and Model Fitting}\label{subsec:Cov-Fitting}

To evaluate the constraints of these probes, we first calculate the covariance matrix by assuming the main contribution is from the Gaussian covariance, and ignore the non-Gaussian terms \citep{Hu&Jain2004}. Then we have
\begin{equation}
\begin{aligned}
    &{\rm Cov}\left[ \Tilde{C}_{\text{XY}}^{ij}(\ell), \Tilde{C}_{\text{XY}}^{mn}(\ell^{\prime}) \right] \\
    &= \frac{\delta_{\ell \ell ^{\prime}}}{f_{\text{sky}}\Delta\ell(2\ell+1)}
    \left[ 
    \Tilde{C}_{\text{XY}}^{im}(\ell) \Tilde{C}_{\text{XY}}^{jn}(\ell) + 
    \Tilde{C}_{\text{XY}}^{in}(\ell) \Tilde{C}_{\text{XY}}^{jm}(\ell) 
    \right],
    \label{eq:cov-1}
\end{aligned}
\end{equation}
where $\text{XY} = \{\text{gg}, \gamma \gamma, \gamma\text{g}\}$, and $f_{\rm sky}$ stands for the sky fraction. For the {\it CSST} survey $f_{\rm sky}$ is about 42\% sky coverage . In our case, we have 30 power spectra in total (after discarding the low-amplitude ones).

Then we fit the mock data of the {\it CSST} photometric surveys by using the $\chi^2$ method, which takes the form as
\begin{equation}
    \chi^2 = \sum_{\ell_{\text{min}}}^{\ell_{\text{max}}}\left[ 
    \vec{d}(\ell) - \vec{t}(\ell) \right] 
    {\rm Cov}^{-1} \left[ \vec{d}(\ell), \vec{d}(\ell^{\prime}) \right]
    \left[ \vec{d}(\ell) - \vec{t}(\ell) \right],
    \label{eq:chi2}
\end{equation}
where $\vec{d}(\ell)$ and $\vec{t}(\ell)$ are the observed and theoretical data vectors, respectively. And the likelihood function takes the form as $\mathcal{L} \propto {\rm exp}(-\chi^2/2)$.

\begin{figure*}
	\includegraphics[scale = 1.4]{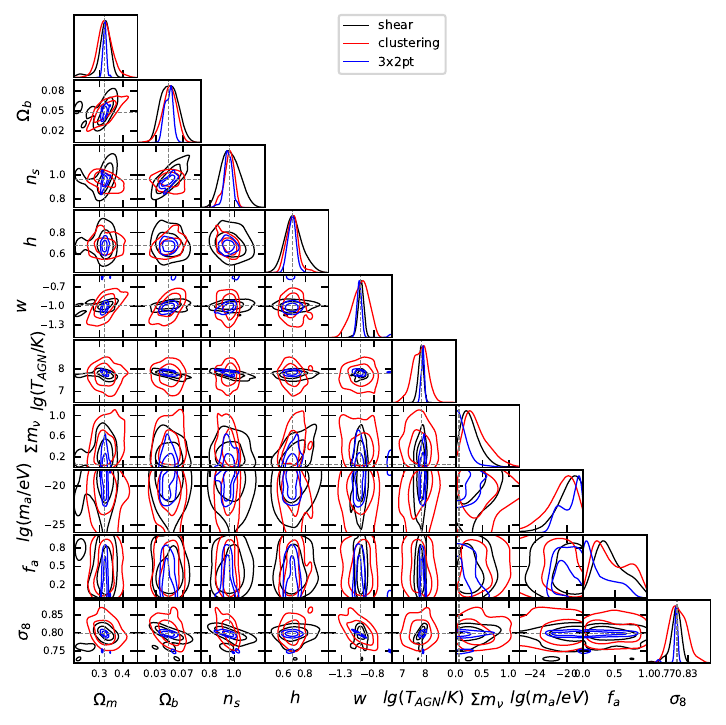}
    \caption{The contour maps of the seven cosmological parameters, one baryonic feedback strength parameter and two axionic parameters with 68\% and 95\% CL for the $\it{CSST}$ galaxy clustering (red), cosmic shear (black), and 3$\times$2pt (blue) surveys. The 1D probability distribution function (PDF) of each parameter is also shown. The grey vertical and horizontal dash lines stand for the fiducial values of these parameters.}
    \label{fig:constraint-1}
\end{figure*}

We sample the posterior distribution of model parameters using the Markov Chain Monte Carlo (MCMC) method by making use of the {\tt emcee} package \citep{emcee}, which is a widely used MCMC Ensemble sampler based on the affine-invariant ensemble sampling algorithm \citep{Goodman2010}. We initialize 50 walkers around our fiducial parameters, and obtain about 500 thousand steps. The first $30\%$ steps are discarded as the burn-in. The free parameters that we include for each survey, and their fiducial values and priors are listed in Table~\ref{tab:params-1}. 

We include dark energy equation of state $w$, reduced Hubble constant $h$, neutrino total mass $\sum m_{\nu}$ and other four cosmological parameters in the constraining process. The baryonic feedback parameter $\text{log}_{10}(T_{\text{AGN}}/\text{K})$ is also considered. For the ULA, we include two parameters, i.e. the ULA dark matter fraction $f_{\text{a}} = \Omega_{\text{a}}/\Omega_{\text{d}}$ and the ULA mass $\text{log}_{10}(m_{\text{a}}/\text{eV})$. In order to explore the ULA effect and constrain the mass and fraction ranges of the ULA, we assume a pure CDM scenario as our fiducial cosmology to generate the mock data.
The systematical parameters, such as the ones from galaxy bias, photo-$z$ uncertainty, intrinsic alignment and shear calibration, are also considered. Totally we have 7 cosmological parameters, 1 baryonic feedback parameter, 2 ULA parameters, and 18 systematical parameters in the fitting process.


\section{Constraint results}\label{sec:Results}


\begin{table*}
    \centering
        \caption{The 68\% CL constraint result of the 7 cosmological parameters, 2 axionic parameters, 1 baryonic parameter and 2 intrinsic alignment parameters in our model for the CSST galaxy clustering, weak lensing, and 3$\times$2pt surveys. The constraint relative accuracy for each parameter is also shown in the bracket.}
    \begin{tabular}{lcccccr}
       \hline
       \text{Parameter} & \text{Fiducial Value} & \text{Constraints by} & \text{Constraints by} & \text{Constraints by} \\
         & & \text{Galaxy Clustering} & \text{Weak Lensing} & \text{3$\times$2pt}\\
       \hline
       \text{Cosmological Parameter} \\
       \hline
       $\Omega_{\text{m}}$ & 0.32 & $0.324^{+0.027}_{-0.033}$ ($9.26\%$) & $0.318^{+0.030}_{-0.016}$ ($7.23\%$) & $0.3219^{+0.0070}_{-0.0079}$ ($2.31\%$) \\
       $\Omega_{\text{b}}$ & 0.048 & $0.0494^{+0.012}_{-0.0093} ($21.56\%$)$ & $0.049^{+0.013}_{-0.013}$ ($26.53\%$) & $0.0482^{+0.0067}_{-0.0058}$ ($12.97\%$) \\
       h & 0.6774 & $0.680^{+0.056}_{-0.056}$ ($8.24\%$) & $0.692^{+0.070}_{-0.090}$ ($11.56\%$) & $0.676^{+0.037}_{-0.037}$ ($5.47\%$) \\
       $n_{\text{s}}$ & 0.96 & $0.955^{+0.050}_{-0.032} $ ($4.29\%$) & $0.968^{+0.071}_{-0.085}$ ($8.05\%$) & $0.949^{+0.033}_{-0.030} $ ($3.32\%$) \\
       $w$ & -1 & $-1.01^{+0.15}_{-0.11}$ ($12.87\%$) & $-1.000^{+0.046}_{-0.039}$ ($4.25\%$) & $-1.001^{+0.014}_{-0.061}$ ($3.75\%$) \\
       $\Sigma m_{\nu}$ & 0.06 eV & $< 0.72 \text{ eV}$ & $< 0.55 \text{ eV}$ & $< 0.29 \text{ eV}$\\
       $\sigma_8$ & 0.8 & $0.804^{+0.022}_{-0.025}$ ($2.92\%$)  & $0.796^{+0.016}_{-0.0094}$ ($1.60\%$) & $0.7957^{+0.0064}_{-0.0017}$ ($0.51\%$) \\
       \hline
        \text{ULA Parameter} \\
       \hline
       $f_{\text{a}}$ & -- & $\leqslant 0.91$ & $\leqslant 0.84$ & $\leqslant 0.68$\\
       $\text{log}_{10}(m_{\text{a}}/\text{eV})$ & -- & $\geqslant -23.4$ & $\geqslant -22.3$ & $\geqslant -21.9$ \\
       \hline
       \text{Baryonic Effect} \\
       \hline
       $\text{log}_{10}(T_{\text{AGN}}/\text{K})$ & 7.8 & $7.72^{+0.34}_{-0.34}$ ($4.40\%$) & $7.75^{+0.12}_{-0.099}$ ($1.41\%$) & $7.831^{+0.070}_{-0.070}$ ($0.89\%$)\\
       \hline
       \text{Intrinsic Alignment} \\
       \hline
       $A_{\text{IA}}$ & 1 & -- & $1.006^{+0.089}_{-0.081}$ ($8.45\%$) & $0.997^{+0.025}_{-0.027}$ ($2.61\%$) \\
       $\eta_{\text{IA}}$ & 0& -- & $-0.09^{+0.38}_{-0.31}$ ($383\%$) & $0.047^{+0.029}_{-0.052}$ ($86.17\%$) \\
       \hline
    \end{tabular}

    \label{tab:params-2}
\end{table*}

We show the constraint result of the whole set of cosmological parameters by each individual tracers in Fig.\ref{fig:constraint-1} with 68\% and 95\% confidence level (CL) contours, and the marginalized 1D probability distribution functions. The grey dash lines mark the fiducial values of the parameters. The best-fit values, 68\% CL and relative accuracies of the parameters are listed in Tab.\ref{tab:params-2}. We can find that, taking the ULAs into account, the CSST photometric survey can constrain the matter density parameter $\Omega_{\text{m}}$ and equation of state of dark energy $w$ to relative accuracies $\lesssim 3\%$ and $\lesssim 4\%$, respectively. We notice that although they are slightly larger ($\sim1\%-2\%$) compared to the results without considering the ULAs \citep{Lin-2022}, which is because we have included more free parameters of the ULA model in the current case, they are basically in the same level. This is due to that the high mass ULAs act like CDM except at small scales, so it would not significant impact the constraint on other standard cosmological parameters when considering the ULAs.

For the neutrino total mass $\Sigma m_{\nu}$, we find that $\Sigma m_{\nu}<1.05$, $0.86$, and $0.74$ eV at 95\% CL for the CSST galaxy angular clustering, weak lensing, and 3$\times$2pt surveys, respectively. Compared to the result when excluding the ULAs, i.e. $<0.56$ eV at 95\% CL for the CSST 3$\times$2pt survey given in \cite{Lin-2022}, it obviously becomes larger. As shown in Fig.~\ref{fig:constraint-1}, we can find a degeneracy between $\Sigma m_{\nu}$ and the ULA mass $m_{\rm a}$ as expected, that $m_{\text{a}}$ will be larger (smaller) when $\Sigma m_{\nu}$ becomes larger (smaller) and vice versa, since large $\Sigma m_{\nu}$ can significantly suppress the power spectra at small scales, which results in a similar effect as light ULAs with small $m_{\rm a}$. This means that the constraint on the neutrino total mass $\Sigma m_{\nu}$ can be significantly affected by including the ULAs, and make the constraint on $\Sigma m_{\nu}$ looser, just as shown in our result.

In Fig.~\ref{fig:constraint-1}, the constraint results of $\text{log}_{10}(m_{\text{a}}/\text{eV})$ and $f_{\text{a}}$ are also given. We can see the constraint power on these two ULA parameters by the CSST galaxy clustering survey are relatively weak, which are $\text{log}_{10}(m_{\text{a}}/\text{eV})\geqslant -23.4$ and -25.9, and $f_{\text{a}}\leqslant0.91$ and 0.97 at 68\% and 95\% CL, respectively. This is because, in the high ULA mass range we are exploring, the suppression on matter power spectrum is only significant at small scales with $k\gtrsim1\ h\,{\rm Mpc^{-1}}$. Since the galaxy angular power spectra at small scales have been cut off to avoid the uncertainty from the non-linear effect, especially for galaxy bias, much information of the ULA is lost. On the other hand, weak lensing can directly probe the dark matter distribution even on small scales ($\ell\sim3000$). Hence more stringent constraints could be obtained, and we have $\text{log}_{10}(m_{\text{a}}/\text{eV})\geqslant-22.3$ and -25.4, and $f_{\text{a}}\leqslant0.84$ and 0.95 at 68\% and 95\% CL. 
When considering the 3$\times$2pt survey, more information can be included, and systematics, e.g. photo-$z$ uncertainties, could be effectively determined, so that we can find tighter constraints on the ULA parameters. Then we obtain
\begin{equation}
\begin{split}
    \hspace{1em} \text{log}_{10}(m_{\text{a}}/\text{eV}) \geqslant -21.9\ {\rm and}\ -22.5 \hspace{1em} (\text{68\% and 95\% CL}), \\
    \hspace{1em} f_{\text{a}} \leqslant 0.68\ {\rm and}\ 0.83 \hspace{1em} (\text{68\% and 95\% CL}).
    \label{eq:constraint}
\end{split}
\end{equation}
This indicates that the CSST 3$\times$2pt photometric survey probably cannot distinguish the CDM and ULA models when $m_{\text{a}} \geqslant 10^{-22.5}$ eV and $f_{\text{a}} \leqslant 0.83$ if considering the baryonic feedback effect. Besides, we can see that the constraint on $f_{\rm a}$ is weak and close to 1 at 95\% CL for all the three datasets, since the current ULA mass bound we obtain is relatively high, which will appear similarly to the CDM in the CSST galaxy clustering and weak lensing surveys.

When comparing our results to previous studies, we notice that they usually do not consider the effect of baryonic feedback and assume that all dark matter are contributed by the ULAs (i.e. assuming $f_{\rm a}=1$). In order to make the comparison, we also derive the constraint results without the baryonic feedback, and the ULA fraction $f_{\rm a}$ is still kept as a free parameter in this case. Then we find
\begin{equation}
\begin{split}
    \hspace{1em} \text{log}_{10}(m_{\text{a}}/\text{eV}) \geqslant -19.9\ {\rm and}\ -21.9 \hspace{1em} (\text{68\% and 95\% CL}), \\
    \hspace{1em} f_{\text{a}} \leqslant 0.40\ {\rm and}\ 0.77 \hspace{1em} (\text{68\% and 95\% CL}).
    \label{eq:constraint}
\end{split}
\end{equation}

\begin{figure}
    \centering
    \includegraphics[scale = 0.6]{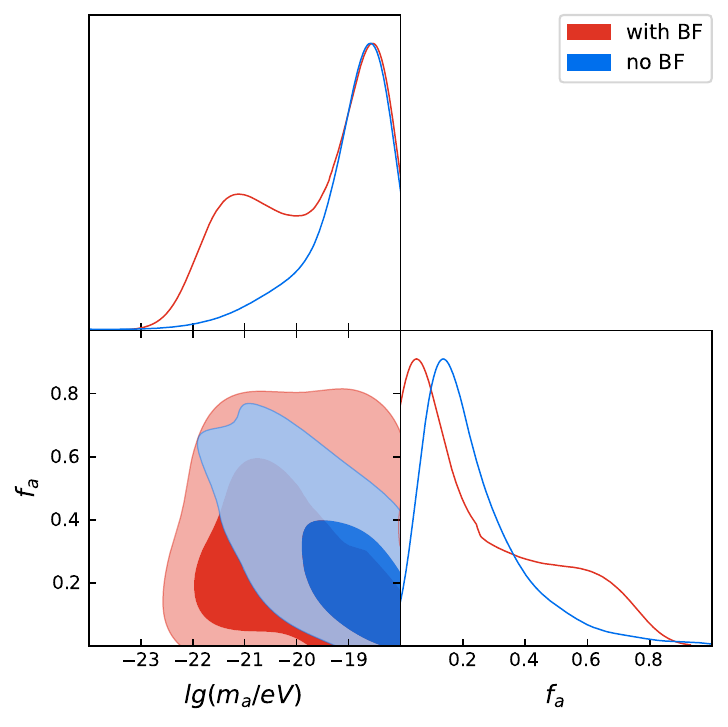}
    \caption{The contour maps and 1D PDFs of $\text{log}_{10}(m_{\text{a}}/\text{eV})$ and $f_{\text{a}}$ with and without the baryonic feedback for the CSST 3$\times$2pt survey are shown in blue and red contours and curves, respectively.}
    \label{fig:BFnoBF}
\end{figure}

The contour maps and 1D PDFs of $\text{log}_{10}(m_{\text{a}}/\text{eV})$ and $f_{\text{a}}$ with and without considering the baryonic feedback for the CSST 3$\times$2pt survey are shown in Figure~\ref{fig:BFnoBF}. In \cite{Dentler-2022}, they obtain a lower limit of $m_{\text{a}} \geqslant 10^{-24.6}$ eV at $95\%$ CL from the DES Y1 shear data, and the result can be further improved by combining DES Y1 shear data with $\it{Planck}$ 2018 data, which gives $m_{\text{a}} \geqslant 10^{-23}$ eV ($95\%$ CL). As can be seen, the CSST 3$\times$2pt survey can improve the constraint on the ULA mass by one order of magnitude at least than the current result. The constraint can be further tightened by including $\it Planck$ data or future CMB observations and other kinds of surveys.
On the other hand, we also compare our result to the predictions of other Stage-IV surveys like $\it Euclid$. For example, \cite{Kunkel-2022} investigates the constraint power using the power spectra, bispectra and trispectra for a $\it Euclid$-like weak lensing survey, and they find a ULA mass lower limit $m_{\text{a}} \geqslant 10^{-22}$ eV (95\% CL) assuming no baryonic feedback and $f_{\rm a}=1$. This is similar to our result without the baryonic feedback effect. 

Based on the discussion above, for the current and next generation Stage-IV photometric galaxy clustering and weak lensing surveys, the constraint on the ULA mass probably cannot be greater than $m_{\rm a} = 10^{-22}-10^{-21}$ eV, especially when considering the effects at small scales like baryonic feedback. As shown in Fig.\ref{fig:matter-ps}, we can find that, the suppression scales of the ULA at high mass range is about $1 \sim 10\, h$ Mpc$^{-1}$. It is still difficult for the current and next-generation weak lensing or galaxy angular clustering surveys to explore such high-$k$ region precisely, and it is also challenging to model the complicated physics and effects accuratly at these non-linear scales.

Nevertheless, if we can use other probes to calibrate the small-scale effects like baryonic feedback, we can indeed obtain better constraints on both the ULA and cosmological parameters. For example, we can adopt the cross correlations of the diffuse X-ray background \citep{Schneider-2020, Ferreira-2023},  thermal Sunyaev-Zeldovich (tSZ) effect \citep{Hojjati-2017, Troster-2022}, and Fast Radio Bursts (FRBs) dispersion measure (DM) \citep{Reischke-2023} with the cosmic shear measurements. These probes are sensitive to the temperature and distribution of free electrons in the Universe, which are tightly related to the properties of gas in dark matter haloes. Hence this kind of measurements probably can help us to accurately investigate the baryonic feedback and other process, and further improve the constraint on the ULA.

\section{Summary}\label{sec:Conclusion}

In this work, we forecast the constraint on the ULA particle mass with the $\it CSST$ weak lensing, galaxy angular clustering, and 3$\times$2pt surveys. We generate the mock data based on the COSMOS catalog to obtain galaxy redshift distribution, surface density, and other necessary information. The neutrinos, baryonic feedback, and systematics from galaxy bias, intrinsic alignment, photo-z uncertainties, shear calibration, and instruments are included in our analysis. We assume a pure CDM scenario as the fiducial cosmology, and employ the MCMC technique in the fitting process.

We find that the CSST photometric surveys can provide precise constraints on the matter energy density paramter $\Omega_{\rm m}$ and the equation of state of dark energy $w$ with relative accuracies higher than $3\%$ and $4\%$, respectively, when including the ULAs. This is similar to the result without the ULAs, and only becomes $\sim 1\%-2\%$ larger. On the other hand, the constraint on the neutrino total mass $\Sigma m_{\nu}$ is obviously looser than that excluding the ULAs. This is due to relatively strong degenarcy between the particle mass of neutrino and ULA in the weak lensing and galaxy clustering surveys. Neutrinos with high particle mass will suppress the matter power spectrum in a similar way as light ULAs, which would significantly affect the constraint on $\Sigma m_{\nu}$ when considering the ULAs.

For constraining the ULA model, we obtain a lower limit for the ULA particle mass $\text{log}_{10}(m_{\text{a}}/\text{eV}) \geqslant -22.5$ and an upper limit for the ULA fraction $f_{\text{a}} \leqslant 0.83$ at 95\% CL when considering the baryonic feedback for the $\it{CSST}$ 3$\times$2pt survey. 
In order to compare to the results derived from the studies of current (e.g. DES Y1) and other Stage-IV (like $\it Euclid$) galaxy surveys, which do not consider the baryonic feedback and assume $f_{\text{a}}=1$, we also investigate the results without the baryonic feedback effect. In this case, we obtain $\text{log}_{10}(m_{\text{a}}/\text{eV}) \geqslant -21.9$ with $f_{\text{a}} \leqslant 0.77$ (95\% CL) when ignoring  the baryonic feedback for the $\it{CSST}$ 3$\times$2pt survey. We find that the CSST photometric galaxy and weak lensing surveys can provide similar constraints as other next-generation surveys, which can improve the constrains by one order of magnitude than the current results.

We also notice that it may be difficult to put stringent constraint on the ULA mass larger than $m_{\rm a}=10^{-22}-10^{-21}$ eV for the next-generation galaxy surveys, especially considering the complex physics and effects at small and non-linear scales.
Since the constraint power of a $\it{CSST}$-like survey mostly comes from small scales with non-linear physical process, if we could have a better understanding on the small-scale physics, e.g. baryonic feedback, we probably can effectively reduce the uncertainties, and obtain more reliable and stringent constraint on the ULAs.

\section{Acknowledgements}

We acknowledge the support of National Key R\&D Program of China grant Nos. 2022YFF0503404, 2020SKA0110402, the CAS Project for Young Scientists in Basic Research (No. YSBR-092), the National Natural Science Foundation of China (NSFC, Grant Nos. 11473044 and 11973047), and the Chinese Academy of Sciences grants QYZDJ-SSW-SLH017, XDB23040100, XDA15020200. This work is also supported by science research grants from the China Manned Space Project with grant Nos. CMS-CSST-2021-B01 and CMS- CSST-2021-A01.

\section{Data Availability}

The data that support the findings of this study are available from the corresponding author, upon reasonable request.



\bibliographystyle{mnras}
\bibliography{ref} 




\appendix

\section{Convergence Test}

\begin{figure}
    \includegraphics[scale=0.55]{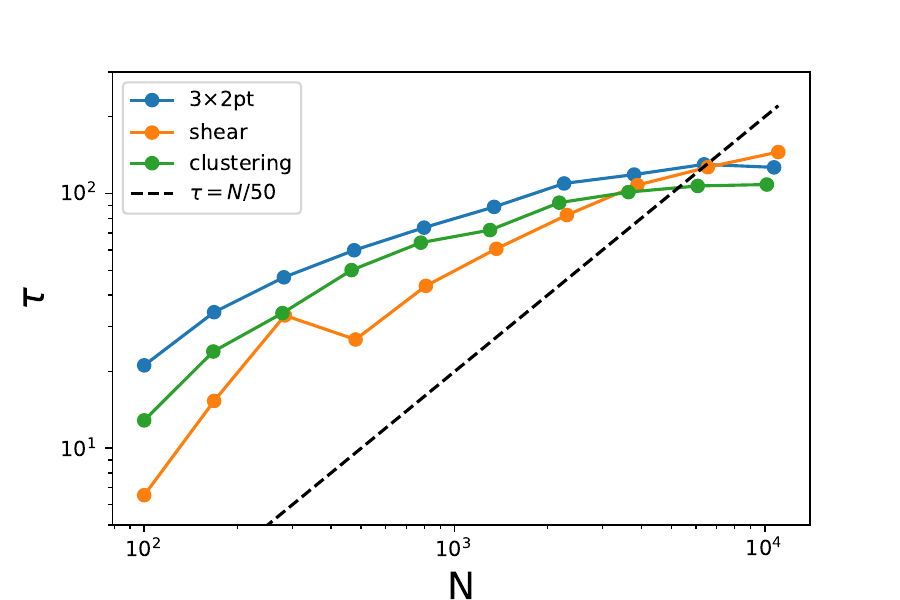}
    \caption{Autocorrelation time estimate for the MCMC chains. $N$ is the number of samples per walker, and $\tau$ is the integrated autocorrelation time. We have confirmed that all chains have reached the convergence threshold $N/\tau > 50$.}
    \label{fig:autocorrelation-time}
\end{figure}

Here we present the convergence test for our MCMC chains. We use the integrated autocorrelation time provided by \cite{Goodman2010}, which is an useful and common method to check the convergence of MCMC chains with multiple walkers.\\

The basic idea of integrated autocorrelation time is that, the samples in the chain are not independent, but we can estimate an effective number of independent samples $N/\tau$ (where $N$ is the number of samples per walker, $\tau$ is the integrated autocorrelation time). If the effective number reaches a threshold, then the chain can be considered as independent. The higher the $N/\tau$ is, the lower sampling errors that we can expect from MCMC, and a typically threshold is $N/\tau > 50$. We present our estimation of autocorrelation time for our three probes in Fig. \ref{fig:autocorrelation-time}, and confirm that all of them have reached the convergence threshold.


\bsp	
\label{lastpage}
\end{document}